# Compressive optical imaging with a photonic lantern

Debaditya Choudhury [1, 2, †], Duncan K. McNicholl [1, 2, †], Audrey Repetti [3, †], Itandehui Gris-Sánchez [4, 5], Tim A. Birks [4], Yves Wiaux [3] and Robert R. Thomson[1, 2, *]

[1] Institute of Photonics and Quantum Sciences, Heriot-Watt University, Edinburgh EH14 4AS, UK

[2] EPSRC IRC Hub, MRC Centre for Inflammation Research, Queen's Medical Research Institute (QMRI), University of Edinburgh, Edinburgh, UK

[3] Institute of Sensors, Signals and System, Heriot-Watt University, Edinburgh EH14 4AS, UK

[4] Department of Physics, University of Bath, Claverton Down, Bath BA2 7AY, UK

[5] Currently with ITEAM Research Institute, Universitat Politècnica de València, Valencia, 46022, Spain

[†]These authors contributed equally to this work.

*R.R.Thomson@hw.ac.uk

**Abstract**

The thin and flexible nature of optical fibres often makes them the ideal technology to view biological processes *in-vivo*, but current microendoscopic approaches are limited in spatial resolution. Here, we demonstrate a new route to high resolution microendoscopy using a multicore fibre (MCF) with an adiabatic multimode-to-single-mode "photonic lantern" transition formed at the distal end by tapering. We show that distinct multimode patterns of light can be projected from the output of the lantern by individually exciting the single-mode MCF cores, and that these patterns are highly stable to fibre movement. This capability is then exploited to demonstrate a form of single-pixel imaging, where a single pixel detector is used to detect the fraction of light transmitted through the object for each multimode pattern. A custom compressive imaging algorithm we call SARA-COIL is used to reconstruct the object using only the pre-measured multimode patterns themselves and the detector signals.

**Introduction**

Endoscopes that use bundles of optical fibres to transmit light in a spatially-selective manner have had a profound impact on minimally-invasive medical procedures. To reduce device size and increase imaging resolution, this concept has been extended to individual fibres containing thousands of light-guiding cores. These single-fibre coherent fibre bundles (SF-CFBs) can provide resolutions of a few microns in the visible [Wood 2018]. When combined with fluorescent contrast agents, they facilitate observation of disease processes at the cellular level [Akram 2018].

SF-CFBs are not without drawbacks. To maintain spatially-selective transmission of light, the fibre cores must be sufficiently spaced to keep core-to-core crosstalk at an acceptable level, intrinsically limiting imaging resolution and throughput. This has led to an explosion of interest in multimode fibre (MMF) imaging, where image information is carried by multiple overlapping spatial modes guided by one multimode core, rather than the many spatially separated cores of the SF-CFB. MMF imaging can deliver an order of magnitude higher spatial resolution, but it is far from trivial to implement because the amplitudes and phases of the MMF modes become scrambled along the fibre. This can be addressed by characterising the MMF's transmission matrix and controlling a spatial light modulator to "undo" the scrambling [Papadopoulos 2012, Čižmár 2012], but any movement of the fibre changes its transmission matrix, and access to the *in-vivo* distal end is required for recalibration unless the new path is precisely known [Plöschner 2015].

Here we demonstrate a new route to high resolution single-fibre microendoscopy using a multicore fibre (MCF) "photonic-lantern" (PL) [Birks 2015]. PLs are guided-wave transitions that efficiently couple light from $N_s$ single mode cores (the MCF) to a multimode waveguide like an MMF. PLs can be made by tapering (heating and stretching in a small flame) a single MCF [Birks 2012], such that the entire reduced-diameter MCF acts as the multimode end of the PL, Fig. 1(a). $N_p = N_s$ distinct multimode patterns of light are generated at the multimode output by coupling light into each core at the MCF input, one at a time. If the MCF exhibits negligible crosstalk between the cores along the length of the MCF, such that the light propagates along just one core, these patterns do not change when bending the fibre, Fig. 1(d-f), unlike those of an ordinary MMF, Fig. 1(g-i). This is because deformation of the MCF merely changes the overall phase of the output pattern. Unlike the spatially-separated modes of a SF-CFB (but like an ordinary MMF), the PL allows the full area of the fibre end-facet to be sampled, and the size of the patterns can be reduced to the minimum allowed by the numerical aperture (NA) of the multimode end.

We demonstrate the feasibility of PL based microendoscopy by using a PL to implement a form of "single-pixel" imaging [Edgar 2019] that we call Compressive Optical Imaging using a Lantern (COIL). Light patterns generated by the PL are projected onto an object (e.g. tissue). Light returned from the object (e.g. fluorescence) is detected by a single-pixel detector, which for the microendoscopy application could be placed at the proximal end of the MCF. The known patterns and measured return signals provide information about the object, from which an image can be formed [Edgar 2019]. We show that the quality and detail of the computed image can be greatly improved by exploiting an advanced image formation algorithm that combines the measurement data with a generic prior postulating that the spatial structure of the image is underpinned by a small number of degrees of freedom. We demonstrate that COIL opens a promising new route to efficient and practical high-resolution microendoscopy.

## Results

### Compressive imaging algorithm

The starting point for our image reconstruction algorithm is to approach PL based imaging in the context of the theory of compressive sampling. In this context, one assumes that the image under scrutiny is sparse in some transform domain linearly related to the pixel domain (e.g. the domain of a wavelet transform [Mallat 09]), that is to say that its spatial structure is underpinned by a small number of degrees of freedom. The sparsity prior information is leveraged to enable image recovery from incomplete data. Compressive sampling approaches have been developed in a wide variety of imaging applications ranging from magnetic resonance imaging [Lustig07, Davies14], and astronomical imaging [Wiaux 2009, Carillo 2012], to ghost imaging [Katz 2009, Sun 2012] and speckle imaging [Kim 2015]. Optimisation algorithms represent the dominant class to solve inverse problems for image recovery from incomplete data. The image estimate is defined as a minimiser of an objective function, consisting of the sum of a data-fidelity term and a sparsity-promoting prior term. The resulting minimisation problem is solved through iterative algorithms progressively minimising the objective function.

We work in a highly compressive sampling regime, *i.e.* for very low ratios of the number of data points (e.g. $N_p$ = 121) to the size of the image formed (e.g. n = 125 × 125 pixels). For COIL, this highly under-sampled regime is of particular interest, as it allows the reconstruction of high resolution images without unrealistic demands on the number of MCF cores. The inverse problem therefore becomes heavily ill-posed and image formation requires strong prior information. With that aim we resort to an advanced "average sparsity" model firstly introduced in astronomical imaging [Carrillo 2012], where multiple wavelet transforms are introduced simultaneously to promote sparsity.

To solve the resulting minimisation problem, we rely on modern "proximal splitting" optimisation algorithms [Combettes 2011, Komodakis 2015] whose main features are a guaranteed fast convergence and low computational complexity. These algorithms have been used in computational imaging in a variety of fields (see [Combettes 2011] and references therein). Building on the "average sparsity" approach we developed a proximal algorithm for COIL, dubbed SARA-COIL (or Sparsity Averaging Reweighted Analysis for COIL). Details of our optimisation approach are provided in the Methods section, together with a description of the associated MATLAB toolbox.

**Experimental techniques and results**

Fig. 1(a) is a schematic of an MCF (with $N_s$ = 25 for clarity) with a PL at one end. For the work reported here, the PL was fabricated at one end of ~3 m of MCF with $N_s$ = 121 single-mode cores in a 11 × 11 square array (Fig. 1(b)) with negligible core-to-core crosstalk at 514 nm. The multimode output end of the PL had a core diameter of ~35 μm and an NA of ~0.22 (Fig. 1(c)). See Methods for fabrication details of the PL. Using computer-controlled alignment, each MCF core could be individually excited using coherent 514 nm laser light, generating $N_p$ = $N_s$ = 121 different multimode patterns of light at the output. Each output pattern was highly stable regardless of the conformation of the MCF, Fig. 1(d-f). This is due to the short length (~ 4 cm) of the PL transition itself and the minimal crosstalk between the MCF cores. In contrast, similar bending of an ordinary MMF changes the output pattern, Fig. 1(g-i).

Our experimental imaging setup is similar to the computational ghost imaging system presented in [Sun 2012], where a spatial light modulator projected random patterns of light onto a test object and detectors measured the fraction of power transmitted through the object. In our experiment the spatial light modulator was replaced with the PL, allowing $N_p = N_s = 121$ different patterns to be projected onto the object by exciting each core of the MCF individually. Initially, we used a simple "knife-edge" as the object. As shown in the object images of Fig. 2, the knife-edge was orientated either horizontally (H) or vertically (V) and positioned to block ~25%, ~50% or ~75% of the pattern projected onto it. As shown in the image panels in Fig. 2, COIL successfully reconstructs images of 125 × 125 pixels using only $N_p$ = 121 patterns. All reconstructions we report using experimental data represent a 0.9 mm × 0.9 mm field of view at the object plane, where the lantern output is imaged with a magnification of ~26 for the purposes of this demonstration. Since the illumination light originates from the lantern itself, the resolution of a near-field imaging modality without the imaging optics would scale by the inverse of the same magnification.

To confirm that COIL is applicable to more complex objects, we repeated the experiment using the objects shown in Fig. 3: an "off-centre cross" and "4 dots" positioned asymmetrically. SARA-COIL can clearly reconstruct the off-centre cross, further confirming the generality of the approach, but cannot reconstruct the small features in the "4-dots" object. To demonstrate how imaging quality might improve by using an MCF PL with more cores, we repeated the data acquisition nine times with the object rotated by 40° between each, acquiring transmission data for each object using effectively $N_p = N_s \times 9 = 1089$ different patterns. As expected, increasing the number of patterns significantly increases image quality for the off-centre cross, Fig. 3. It also reconstructs some features of the "4 dots" object, but falls short of fully resolving them.

To establish that our experimental results are in line with those expected from theory, we also performed reconstructions using simulated data. To simulate the intensity patterns from an ideal $N_s = 121$ PL, we first calculated the field distributions of the 121 lowest-order spatial modes of a circular ideal-mirror waveguide. We then generated a set of 121 mutually-orthonormal but otherwise random coherent superpositions of the modes, and formed intensity patterns by taking the square modulus. The imaging experiment was simulated by computing the overlap integral between each intensity pattern and the object. The intensity patterns and overlap data were then processed using SARA-COIL to reconstruct an image. The simulated reconstructions for both objects, using either $N_p = 121$ (not rotated) or $N_p = 1089$ (9 rotations), are shown in Fig. 3 alongside the reconstructions based on experimental data for comparison. As expected, images obtained using both measured and simulated data improved considerably as the number of patterns is increased. Furthermore, if we consider that the multimode port of the PL used in our experiments has a diameter of 35 μm, our $N_p = 1089$ simulations suggest that sub-micron resolution would be achievable using a PL generating only a thousand patterns. (The NA of the port would have to be ~0.3 to support this number of modes, rather than the 0.22 of the PL used here.) Clearly, however, there is a significant difference between the experimental and simulated results. As we discuss later, we believe this is primarily due to limitations with the current experimental setup.

To further highlight the potential of COIL for the high-resolution imaging of structures *in-vivo*, we simulated (as above) the results that might be expected using a $N_s$ = 2000 PL to project $N_p$ = 2000 patterns. The two objects used for this simulation were an image of the 1951 USAF resolution target and a confocal microscope fluorescence image of fixed calcein-stained adenocarcinomic human alveolar basal epithelial (A549) cells. Our images, shown in Fig. 4, are high-quality reconstructions of both objects. Fig. 4 also shows that our image reconstruction technique is robust to the presence of additive Gaussian noise in the overlap data. For example, both contrast and resolution are only minimally affected by the noise, and features such as the horizontal and vertical bars in the top right of the USAF target are still clearly resolvable.

For completeness, Fig. 5 compares SARA-COIL to a simpler, more intuitive, reconstruction algorithm used for classical ghost imaging - see Equation 5 in [Sun 2012]. This algorithm uses only the fractional transmission of the projected pattern to weight its contribution to the image reconstruction. No attempt is made to optimise this towards a realistic object using a prior. The comparison confirms that SARA-COIL significantly improves both resolution and contrast, revealing features that are otherwise barely or not visible. These results provide a compelling justification for the advanced algorithmic approach we adopted.

**Discussion**

The reconstructions presented in Fig. 3 using 1,089 patterns clearly indicate that although our experimental results broadly agree with simulations from ideal data, there is considerable potential for more accurate reconstructions. We highlight that the quality of the reconstructions using experimental data is degraded by the fact that the re-centring of the object onto the pattern after each rotation was only performed by eye, using a thin ring around the object to guide alignment. In fact, both reconstructions in Fig. 3 show hints of resolving this ring. This practical limitation can be readily resolved by adopting MCFs with more cores and not rotating them.

Remarkably, Fig. 4 demonstrates that, even in the presence of noise, an $N_s = N_p = 2000$ COIL system could resolve objects separated by just ~1.6% of the multimode core diameter (see the three-bar pattern at the top right of the USAF target). To put this into context, if a COIL system is constructed to operate using 488 nm excitation light and an $N_s = 2000$ MCF, the multimode output of the PL could have a 63 $\mu$m diameter core with an NA of 0.22, assuming established fabrication techniques [Birks 2012] with an F-doped silica cladding – see Methods. Such a system could resolve objects separated by just ~1.25 $\mu$m. This is close to the 1.35 $\mu$m expected from Rayleigh's criterion (0.61 $\lambda$ / NA), a strong indication that COIL can deliver at least diffraction-limited imaging across the field of view of the core.

The $N_s = N_p$ pattern projection is only the simplest imaging modality one might consider using PLs for. In fact, PLs could enable significantly more advanced and powerful modalities, some driven by compressive sampling principles, but these require the controlled simultaneous excitation of multiple MCF cores to generate coherent combinations of the multimode states at the output. To do this in a controlled manner, the key information to be obtained are the relative phases and amplitudes of the individual basis patterns at the multimode output. We envisage future COIL imaging systems exploiting polarisation maintaining MCFs, where the PL's output is coated to partially reflect some pump light back along the MCF. Since each multimode pattern generates a specific non-binary phase and amplitude distribution across the MCF cores after reflection, and since there is negligible crosstalk between the MCF's cores, the distribution of reflected light across the cores at the proximal end will encode the relative phases and amplitudes of the multimode patterns at the output. In principle, this could facilitate the coherent synthesis of arbitrary excitation fields at the output of the lantern for both near- and far-field spot-scanning modalities, and also enable the projection of many more than $N_s$ different known multimode patterns. As detailed by Mahalati *et al* [Mahalati 2013], the number of possible "intensity modes", and therefore the number of resolvable features across the output core, could reach a maximum of *$4N_s$*. For the case of an $N_s$ = 2000 PL with a 63 μm diameter 0.22 NA multimode core operating at 488 nm, such an approach could deliver a resolution of ~626 nm – significantly smaller than the Rayleigh limit and opening a potential route to super-resolution microendoscopy. The NA of the PL's multimode output can also be pushed well beyond 0.22 by exploiting more advanced fibre approaches. For example, we foresee the creation of PL's using a polarisation maintaining MCF with a double-cladding geometry, such as those commonly used in fibre lasers for efficient cladding pumping. In this case, the MCF cores and their glass cladding would be surrounded by an air cladding that could facilitate a PL multimode port at the distal end with an *in-vivo* NA of up to ~0.65 at 488 nm [Wadsworth 2004]. This might deliver a spatial resolution of ~212 nm, although stability issues during *in-vivo* exposure will obviously play a role in determining this.

We resorted to a powerful framework of optimisation to develop the SARA-COIL algorithm, but further developments may significantly improve image estimation. Firstly, regularisation priors specifically developed for images of interest in microendoscopy can improve quality over our state-of-the-art "average sparsity" prior. Secondly, parallelised "proximal algorithms" [Pesquet 2015, Chambolle 2018] can improve scalability to high-resolution imaging, ultimately to provide real-time microendoscopic imaging. Finally, approximation in the measurement model can severely affect imaging quality in computational imaging (*e.g.* the alignment between object and patterns). Joint calibration and imaging algorithms can be defined in the theory of optimisation, that can simultaneously solve for unknown parameters in the measurement model and form the image [Bolte 2014, Chouzenoux 2016].

**Conclusions**

We have experimentally demonstrated a new form of single-pixel imaging using a multicore fibre and photonic lantern to generate distinct multimode light patterns. We have provided compelling evidence that this, underpinned by the powerful SARA-COIL optimisation algorithm, can deliver at least diffraction-limited imaging across the full area of a multimode fibre core, without sensitivity to bending or any need to control or compensate for modal phases. This meets the world-wide need to develop new fibre-optic imaging techniques to deliver high-resolution images of cellular and molecular mechanisms *in vivo*. We have also discussed how it opens a route to more complex imaging modalities, such as super-resolution microendoscopy with sub-micron resolution. We also anticipate that COIL could also be useful in applications that benefit from a reduced number of measurements, such as fibre-optic epifluorescence or confocal microendoscopy, which are vulnerable to detrimental effects such as photobleaching and phototoxicity [Flusberg 2005].

**Methods**

**The multicore fibre**

The $N_s$ = 121 square-array multicore fibre was originally fabricated for a study of wavelength-to-time mapping [Chandrasekharan 2016]. The cores were positioned on a square grid with a core-to-core spacing of ~10.53 µm. The mode field diameters of the MCF cores were measured at 514 nm using calibrated near-field imaging. The 1/e$^2$ mode field diameter was ~2.1 ± 0.2 µm.

**Photonic lantern fabrication**

To fabricate the PL [Chandrasekharan 2016], the MCF was threaded into a fluorine-doped silica capillary, the refractive index of which is lower than the pure silica cladding of the MCF. The capillary was collapsed, by surface tension, on top of the MCF using an oxybutane flame. Using a similar flame, the cladded structure was then softened and stretched by a tapering rig, forming a biconical fibre-like structure. The multimode port of the PL was finally revealed by cleaving the centre of the tapered waist. The resultant multicore-to-multimode taper was ~4 cm long, with an approximately linear profile. The multimode port's core diameter was ~35 µm and its numerical aperture was 0.22.

**SARA-COIL algorithm**

The observed data, denoted by $y \in \mathbb{R}^{N_p}$ (there is one data point per pattern), consist of a linear transform of the image of interest $x \in \mathbb{R}^n$ with a linear operator whose lines consist of the projection patterns. The measurement model thus reads:

$$y = \Phi x + e,$$

where $\Phi \in \mathbb{R}^{N_p \times n}$ represents the measurement operator and $e \in \mathbb{R}^{N_p}$ the acquisition noise.

The SARA-COIL algorithm results from an adaptation of the "Sparsity Averaging Reweighted Analysis" approach developed by Carrillo *et al.* [Carrillo 2012]. On the one hand, the minimisation problem solved reads as

$$\text{minimise } \|\Omega\Psi x\|_1 \text{ subject to } x \in [0, +\infty[^n \text{ and } \|y - \Phi x\|_2 \leq \epsilon.$$

The first element in this expression is the sparsity-promoting prior term to be minimised. $\|.\|_1$ denotes the non-differentiable $\ell_1$ norm, traditionally invoked in the context of compressive sampling. $\Psi \in \mathbb{R}^{L \times n}$ is the linear operator defining the sparsity transform, built as the concatenation of 9 wavelet transforms ($L = 9n$) as in Carrillo *et al.* [Carrillo 2012]. $\Omega \in \mathbb{R}^{L \times L}$ is a diagonal weighting matrix computed using a re-weighting procedure introduced by Candès *et al.* [Candès 2008b]. The second element of the expression " $x \in [0, +\infty[^n$ " is a prior term imposing the physical constraint of positivity of the intensity image to be formed. The third element "$\|y - \Phi x\|_2 \leq \epsilon$" is the data-fidelity term imposing that the discrepancy between data and model is bounded by the noise energy $\epsilon$.

To solve this minimisation problem, we developed an iterative algorithm based on the primal-dual forward-backward "proximal algorithm" [Condat 2013, Vu 2013].

**Data Availability**

Raw data will be made available through the Heriot-Watt University PURE research data management system.

**Code Availability**

A MATLAB toolbox gathering the algorithm implementation as well as the data necessary to reproduce our simulations results using the USAF resolution target is available on GitHub at https://basp-group.github.io/SARA-COIL/

**Acknowledgements**

This work was funded through the "Proteus" Engineering and Physical Sciences Research Council (EPSRC) Interdisciplinary Research Collaboration (IRC) (EP/K03197X/1), and by the Science and Technology Facilities Council (STFC) through STFC-CLASP grants ST/K006509/1 and ST/K006460/1, and through STFC Consortium grants ST/N000625/1 and ST/N000544/1. The authors thank Philip Emanuel for the use of his confocal image of A549 cells and Eckhardt Optics for their image of the USAF 1951 target.

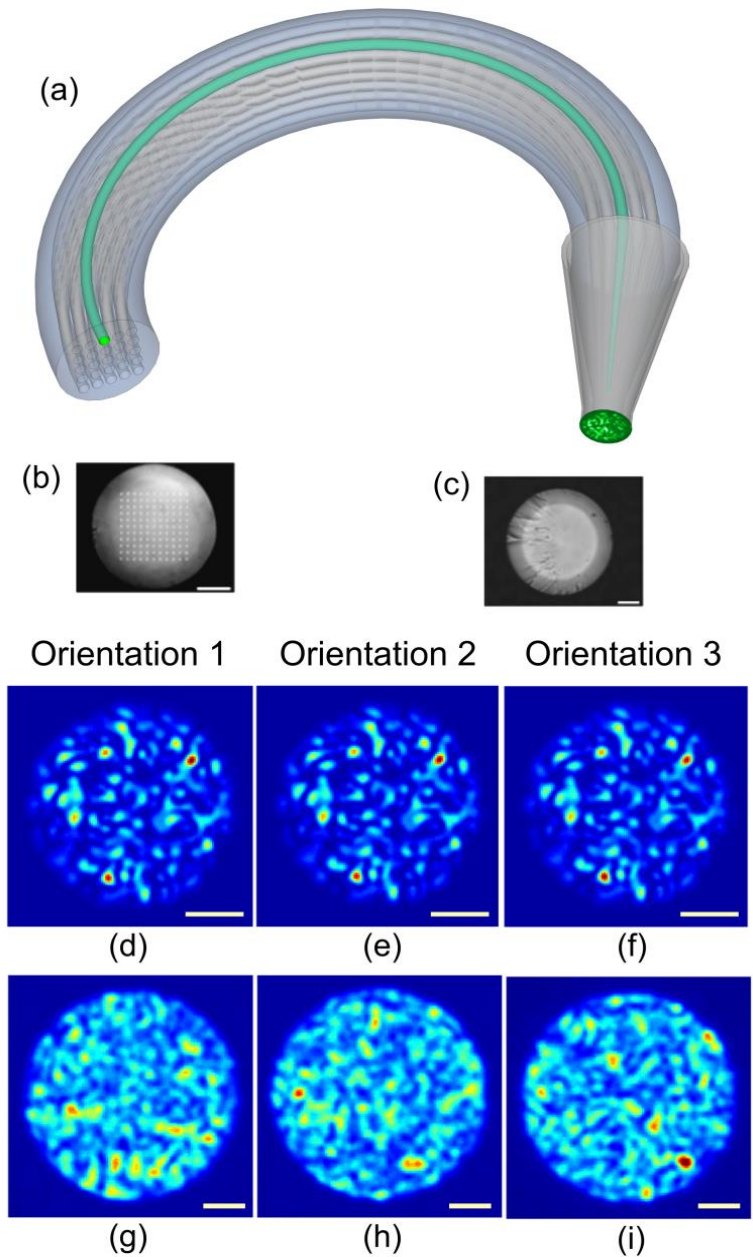

**Fig. 1** (a) Schematic $N_s$ = 25 square-array multicore fibre with a photonic lantern at one end. (in green) Light in one core excites a fixed light pattern at the lantern's output. (b) Optical micrograph of the facet of the $N_s$ = 121 multicore fibre used in this work. Scale bar: 50 µm. (c) Optical micrograph of the multimode output of the photonic lantern. Scale bar: 10 µm. (d - f) Near field intensity patterns at the output of the photonic lantern when one core of the multicore fibre is excited with monochromatic light ($\lambda$ = 514 nm). The patterns are insensitive to fibre bending as shown by the micrographs obtained for three arbitrary conformations of the fibre. Scale bars: 10 µm (g - i) Corresponding near field intensity patterns at the output of a 105 µm core multimode fibre when excited with monochromatic light ($\lambda$ = 514 nm). As shown in the micrographs obtained for three arbitrary conformations of the fibre, the patterns are highly sensitive to bending of the fibre. Scale bars: 20 µm.

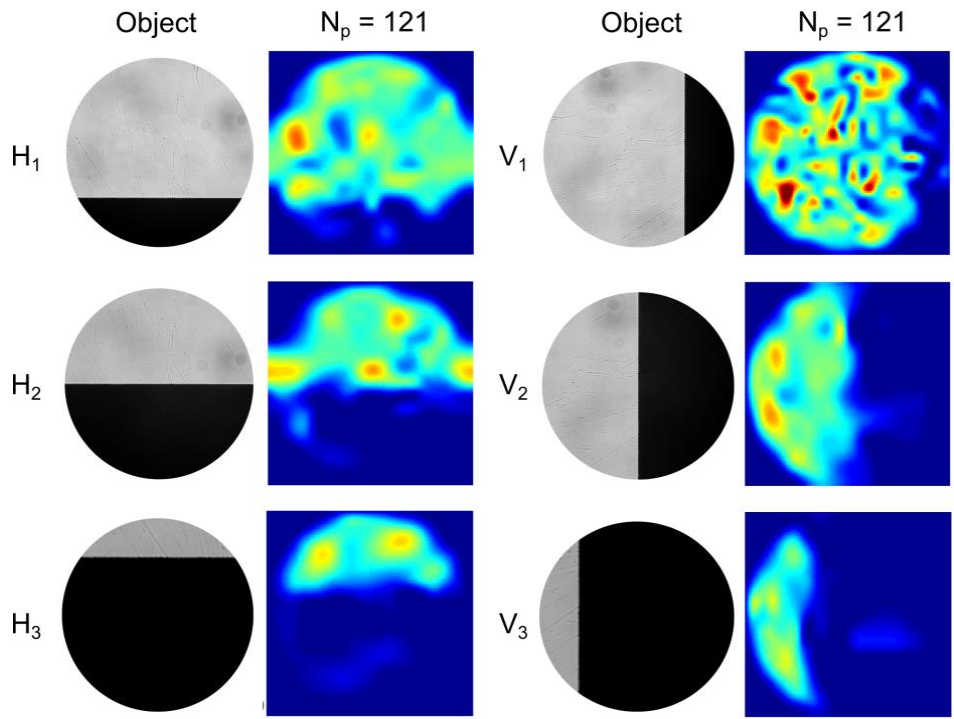

**Fig. 2**: SARA-COIL results obtained using $N_p$ = 121 patterns. Micrographs of the objects are shown in the left of each panel. $H_i$ & $V_i$ respectively denote objects formed by horizontally and vertically overlaying a knife edge over ~25% (i = 1), ~50% (i = 2) and ~75% (i = 3) of the intensity pattern. Each reconstructed image has 125 × 125 pixels, with a field of view in the object plane of 0.9 mm × 0.9 mm.

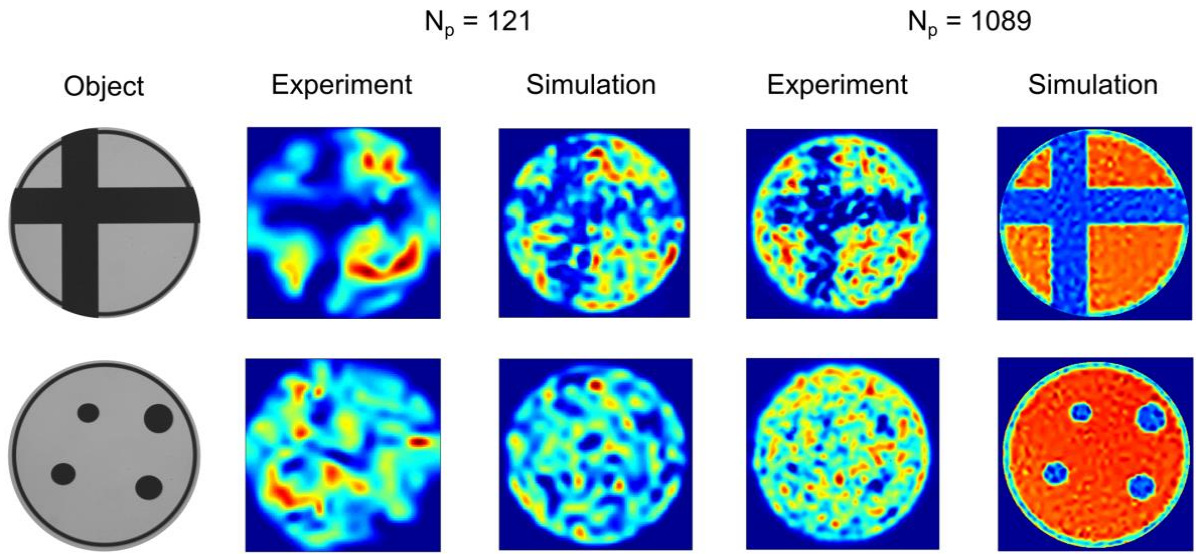

**Fig. 3:** SARA-COIL reconstructions using $N_p$ = 121 and $N_p$ = 1089 patterns, and either measured or simulated patterns and overlap data. The objects are an off-centre cross and 4 asymmetrically-positioned elliptical dots, micrographs of which are presented. For $N_p$ = 1089, the object was rotated about the optical axis by 320° in steps of 40°, effectively creating a total of 121 × 9 patterns. The simulated reconstructions (125 × 125 pixels) for $N_p$ = 121 used patterns generated from random orthonormal superpositions of the 121 lowest-order modes of a circular ideal-mirror waveguide. For $N_p$ = 1089 (377 × 377 pixels) the object was rotated about the optical axis by 320° in steps of 40°. The field of view of all reconstructions is 0.9 mm × 0.9 mm in the object plane.

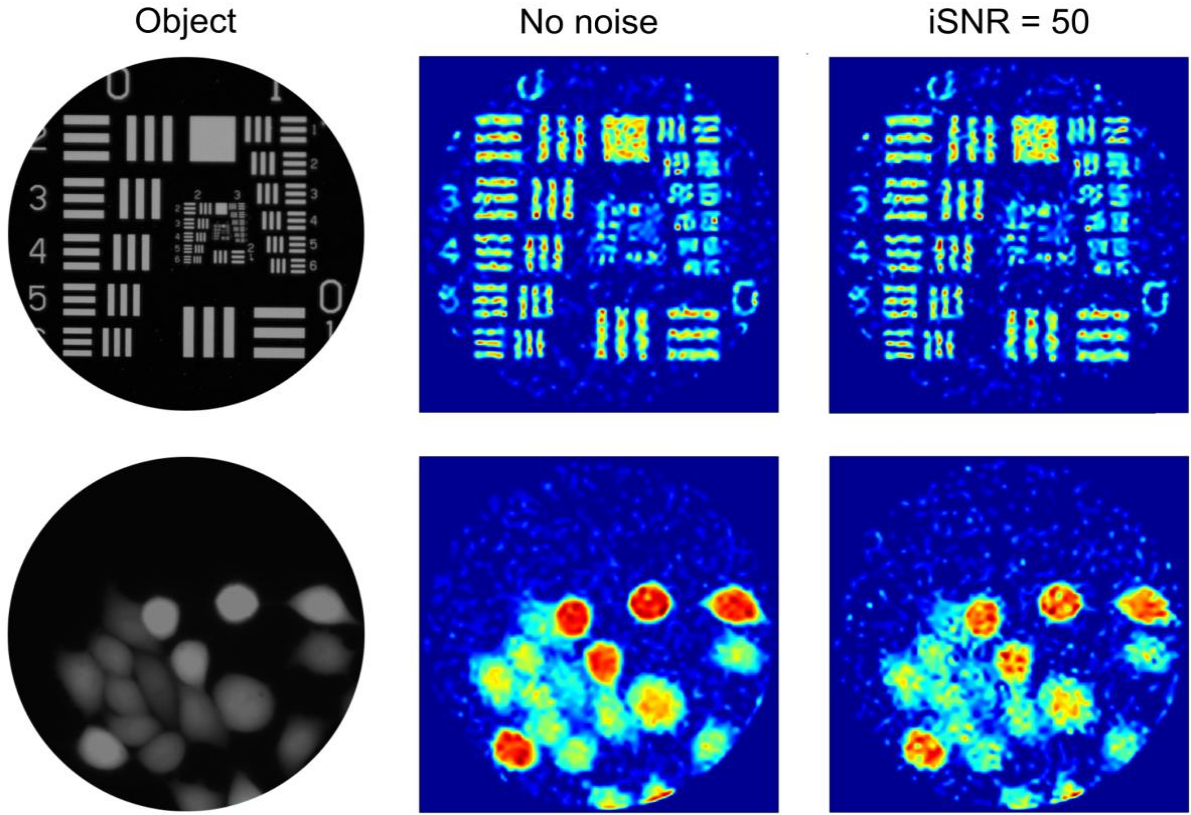

**Fig. 4** Simulated reconstruction results (511 × 511 pixels) obtained using $N_p$ = 2000 intensity patterns generated from random orthonormal superpositions of the 2000 lowest-order modes of a circular ideal-mirror waveguide. The objects were the 1951 USAF resolution target and a confocal microscope image of fixed calcein stained adenocarcinomic human alveolar basal epithelial (A549) cells. For each object, the reconstructed image with additive Gaussian noise (input signal-to-noise ratio iSNR=50) is shown alongside that with no added noise. We highlight the fact that there is deliberately no spatial scale for the reconstructions, since the size of a waveguide supporting $N_p$ = 2000 modes varies depending on its core-cladding refractive index contrast. The reader is referred to the discussion section for more information.

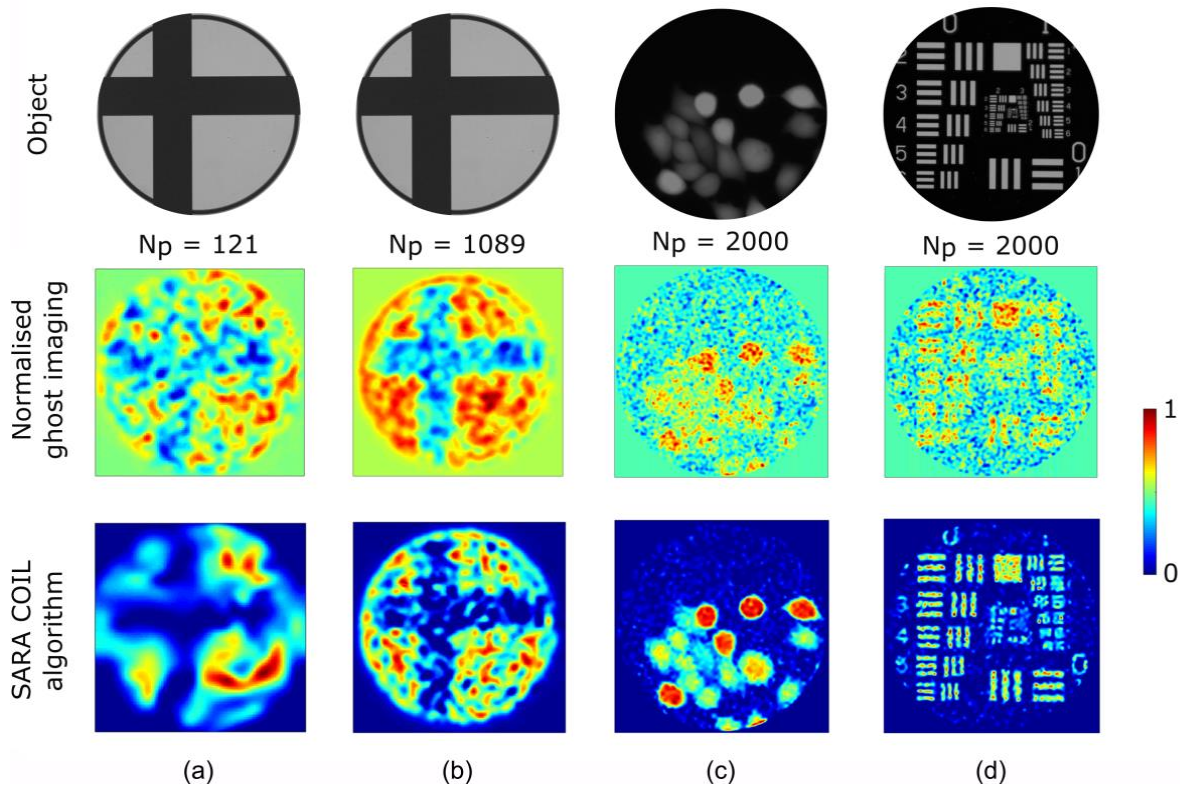

**Fig. 5** Reconstructions of various objects using experimental or simulated data and either an established ghost imaging algorithm (Eq. 5 in [Sun 2012]) (middle row) or SARA-COIL (bottom row). (Column a) 125 × 125 pixel reconstructions of an off-centre cross for $N_p$ = 121 using experimental data. (Column b) 377 × 377 pixel reconstructions of an offset cross for $N_p$ = 1089 using experimental data (Columns c and d) 511 × 511 pixel reconstructions of the A549 cells (c) and the USAF target (d) for $N_p$ = 2000 using simulated patterns and overlap data. Note that regions with no available information are treated differently by the two algorithms. As seen in the corners of all images, the ghost imaging algorithm assigns a mid-scale value, whereas SARA-COIL assigns a value of 0. In images reconstructed from experimental data, 1 represents the regions of highest transmission, and in those based on simulated data 1 represents regions of highest intensity. The field of view of all reconstructions using experimental data is 0.9 mm × 0.9 mm in the object plane.